\documentclass[aps,prb,preprint,superscriptaddress]{revtex4-1}

\usepackage{amsmath}    % need for subequations
\usepackage{graphicx}   % need for figures
\usepackage{color}      % use if color is used in text
\usepackage{subfigure}  % use for side-by-side figures
\usepackage{todonotes}
\usepackage{nicefrac}
\usepackage[latin1]{inputenc}

\usepackage{hyperref}   % use for hypertext links, including those to external documents and URLs

\begin{document}

\title{Electronic structure of undoped and potassium doped coronene investigated by electron energy-loss spectroscopy}
\author{Friedrich Roth}
\author{Johannes Bauer}
\author{Benjamin Mahns}
\author{Bernd B\"uchner}
\author{Martin Knupfer}
\affiliation{IFW Dresden, P.O. Box 270116, D-01171 Dresden, Germany}
\date{\today}

\begin{abstract}
We performed electron energy-loss spectroscopy studies in transmission in order to obtain insight into the electronic properties of potassium intercalated coronene, a recently discovered superconductor with a rather high transition temperature of about 15\,K. A comparison of the loss function of undoped and potassium intercalated coronene shows the appearance of several new peaks in the optical gap upon potassium addition. Furthermore, our core level excitation data clearly signal filling of the conduction bands with electrons.

\end{abstract}

\maketitle

\section{Introduction}

Organic molecular crystals---built from $\pi$ conjugated molecules---have been subject of intense research for a number of reasons.\cite{Granstrom1998,Dodabalapur1995,Tsivgoulis1997,Kaji2009} Due to their relatively open crystal structure their electronic properties can be easily modified by the addition of electron acceptors and donors, which can lead to novel and, in some cases, intriguing or unexpected physical properties. A prominent example for the latter is the formation of metallic, superconducting or insulating phases in the alkali metal doped fullerides depending on their stoichiometry.\cite{Gunnarsson2004,Weaver1994,Knupfer2001,Gunnarsson1997} In particular the superconducting fullerides have attracted a lot of attention, and rather high transition temperatures ($T_c$'s) in e.\,g. K$_3$C$_{60}$ ($T_c$\,=\,18\,K)\cite{Hebard1991}, Cs$_2$RbC$_{60}$ ($T_c$\,=\,33\,K)\cite{Tanigaki1991} or Cs$_3$C$_{60}$ ($T_c$\,=\,38\,K\cite{Ganin2008,Ganin2010} and $T_c$\,=\,40\,K under 15\,kbar \cite{Palstra1995}) have been reported. In this context, further interesting phenomena were reported in alkali metal doped molecular materials such as the observation of an insulator-metal-insulator transition in alkali doped phthalocyanines \cite{Craciun2006}, a transition from a Luttinger to a Fermi liquid in potassium doped carbon nanotubes \cite{Rauf2004}, or the formation of a Mott state in potassium intercalated pentacene. \cite{Craciun2009}

However, in the case of organic superconductors, no new systems with high $T_c$'s similar to those of the fullerides have been discovered in the past decade. Recently, the field was renewed with the discovery of superconductivity in alkali doped picene with a $T_c$ up to 18\,K.\cite{Mitsuhashi2010} Furthermore, after this discovery superconductivity was also reported in other alkali metal intercalated polycyclic aromatic hydrocarbons, such as phenanthrene ($T_c$ = 5\,K) \cite{Wang2011,Andres2011} and coronene ($T_c$ = 15\,K).\cite{Kubozono2011} Motivated by these discoveries and numerous publications on picene, both experimental\cite{Mitsuhashi2010,Roth2010,Roth2011_2,Okazaki2010} and theoretical\cite{Cudazzo2011,Kato2011,Subedi2011,Kim2011,Giovannetti2011,Andres2011_2,Casula2011}, the investigation of the physical properties of coronene in the undoped and doped state is required in order to develop an understanding of the superconducting and normal state properties.

\par

\begin{figure}[ht]
\centering
\includegraphics[width=0.49\linewidth]{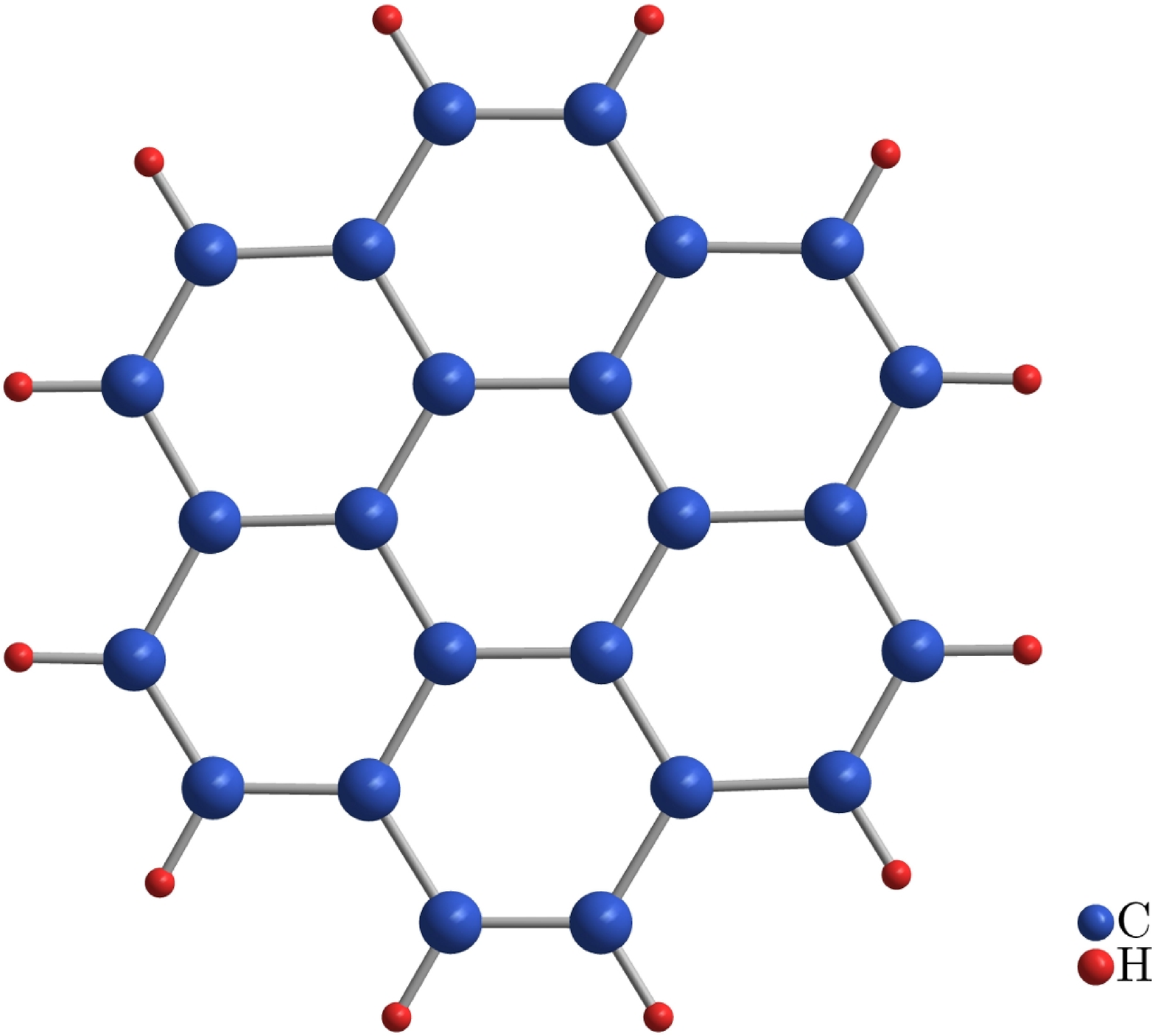}
\includegraphics[width=0.49\linewidth]{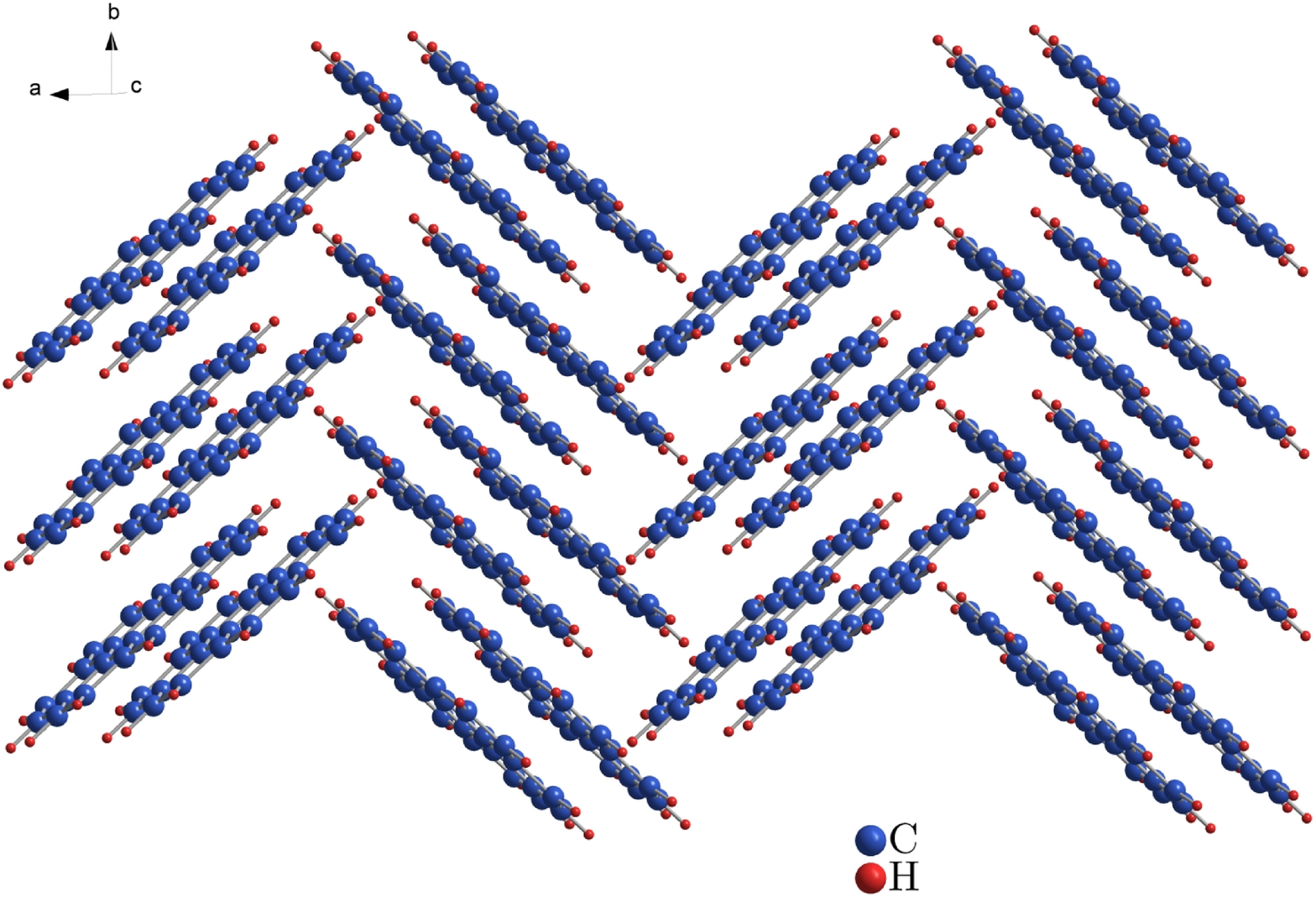}
\caption{Left panel: Schematic representation of the molecular structure of coronene. Right panel: Crystal structure of coronene.} \label{f1}
\end{figure}

The coronene molecule is made out of six benzene rings which are arranged in a circle as depicted in Fig.\,\ref{f1} left panel. In the condensed phase, coronene adopts a monoclinic crystal structure, with lattice constants $a$ = 16.094\,\AA, $b$ =  4.690\,\AA, $c$ = 10.049 \,\AA, and $\beta$ = 110.79$^\circ$, the space group is P2$_1$/a, and the unit cell contains two inequivalent molecules.\cite{Echigo2007} The molecules arrange in a herringbone manner (cf. Fig.\,\ref{f1}) which is typical for many aromatic molecular solids. Furthermore, coronene crystals show two structural phase transitions depending on pressure and temperature in the range between 140---180\,K \cite{Yamamoto1994,Totoki2005} and at 50\,K.\cite{Nakatani1994}

\par

In this contribution we report on an investigation of the electronic properties of undoped and potassium doped coronene using electron energy-loss spectroscopy (EELS). EELS studies of other undoped and doped molecular materials in the past have provided useful insight into their electronic properties.\cite{Schuster2007,Roth2010_2,Knupfer1999_2} We demonstrate that potassium addition leads to a filling of the conduction bands and the appearance of several new low energy excitation. In addition, our analysis of the spectra allows a determination of the dielectric function of K doped coronene.

\section{Experimental}

\noindent Thin films of coronene were prepared by thermal evaporation under high vacuum onto single crystalline KBr substrates kept at room
temperature with a deposition rate of 0.8\,nm/min and a evaporation temperature of about 500\,K. The film thickness was about 100\,nm. These coronene films were floated off in destilled water, mounted onto standard electron microscopy grids and transferred into the spectrometer. Prior to the EELS measurements the films were characterized \textit{in-situ} using electron diffraction. All observed diffraction peaks were consistent with the crystal structure of coronene.\cite{Echigo2007,Kubozono2011} Moreover, the diffraction spectra show no significant pronounced texture which leads to the conclusion that our films are essentially polycrystalline.
                                                    
All electron diffraction studies and loss function measurements were carried out using the 172\,keV spectrometer described in detail elsewhere.\cite{Fink1989} We note that at this high primary beam energy only singlet excitations are possible. The energy and momentum resolution were chosen to be 85\,meV and 0.03\,\AA$^{-1}$, respectively. We have measured the loss function Im[-1/$\epsilon(\textbf{q},\omega)$], which is proportional to the dynamic structure factor S($\textbf{q},\omega$), for a momentum transfer $\textbf{q}$ parallel to the film surface [$\epsilon(\textbf{q},\omega)$ is the dielectric function]. The C\,$1s$ and K\,$2p$ core level studies were measured with an energy resolution of about 200\,meV and a momentum resolution of 0.03\,\AA. In order to obtain a direction independent core level excitation information, we have determined the core level data for three different momentum directions such that the sum of these spectra represent an averaged polycrystalline sample.\cite{Egerton1996} The core excitation spectra have been corrected for a linear background, which has been determined by a linear fit of the data 10\,eV below the excitation threshold.

Potassium was added in several steps by evaporation from a commercial SAES (SAES GETTERS S.p.A., Italy) getter source under ultra-high vacuum
conditions (base pressure lower than 10$^{-10}$\,mbar) until a doping level of about K$_3$coronene was achieved, which is reported to be the superconducting phase.\cite{Kubozono2011} In detail, in each doping step, the sample was exposed to potassium for 5\,min, the current through the SAES getter source was 6\,A and the distance to the sample was about 30\,mm.
During potassium addition, the film was kept at room temperature. 

In order to perfom a Kramers-Kronig analysis (KKA), the raw date have been corrected by substracting contributions of multiple scattering
processes and elimination of contributions of the direct beam by fitting the plasmon peak with a model function.\cite{Schuster2009}

\section{Results and discussion}

\begin{figure}[h]
\centering
\includegraphics[height=5.8cm]{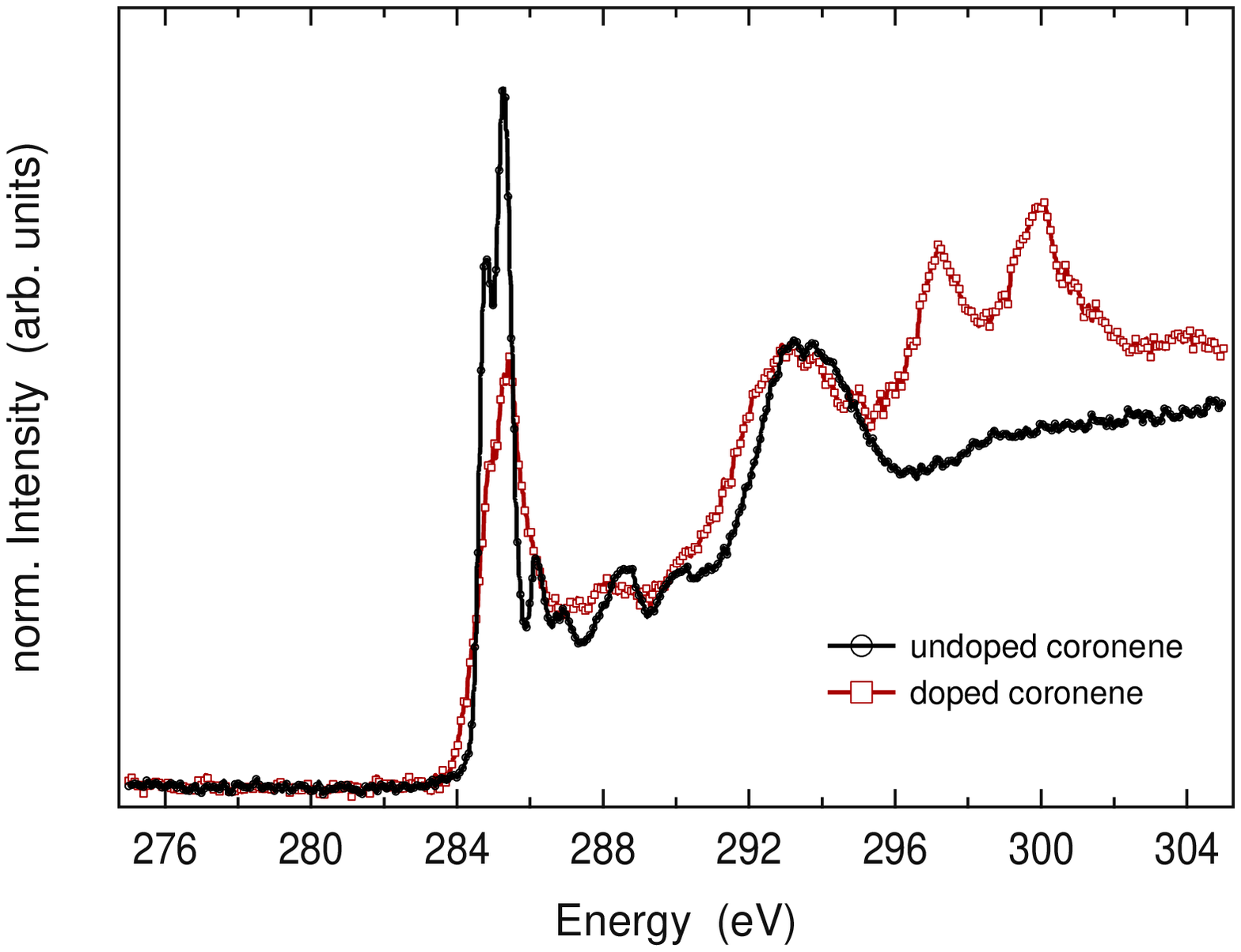}
\includegraphics[height=5.8cm]{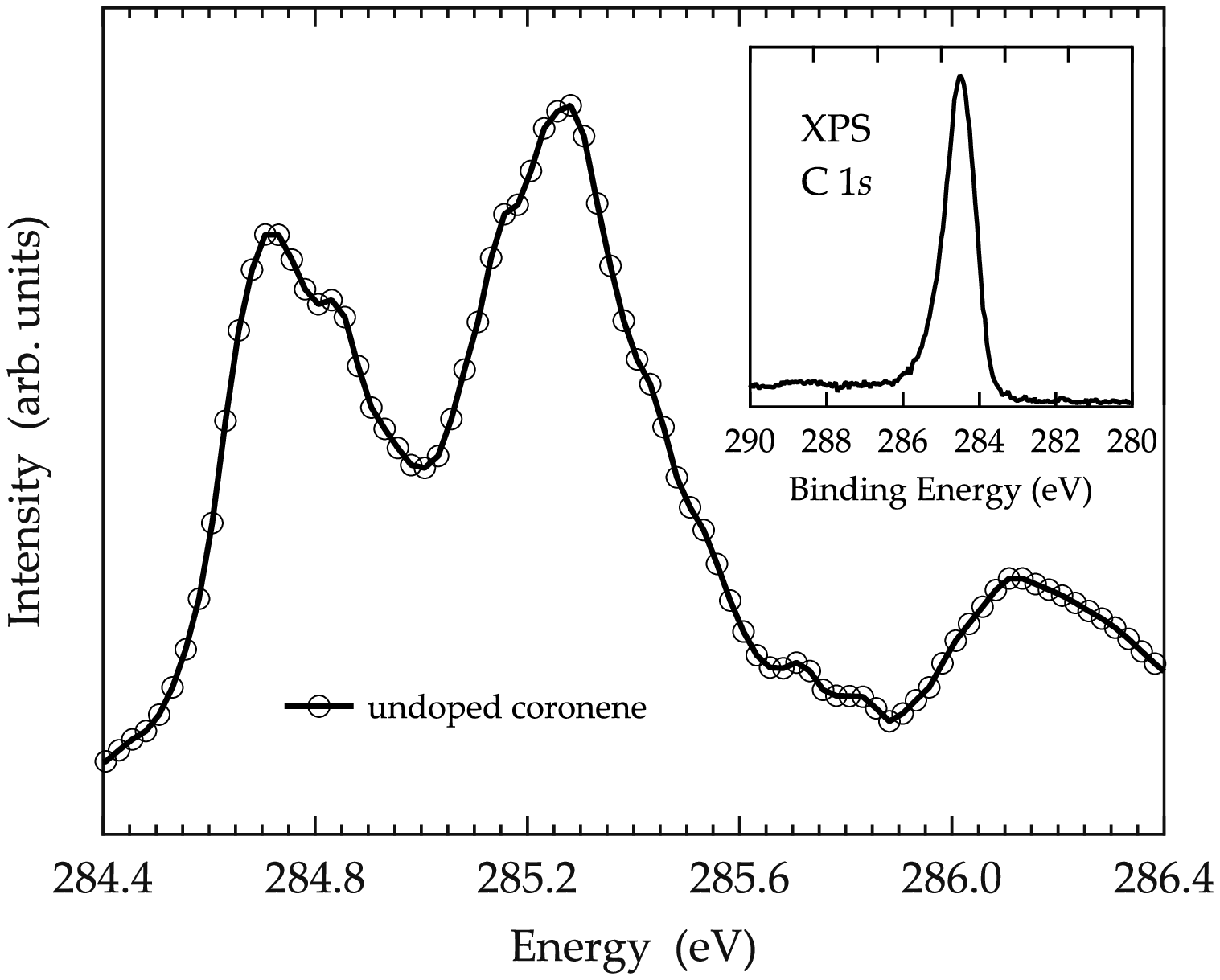}
\caption{Left panel: C\,$1s$ and K\,$2p$ core level excitations of undoped (black circles) and potassium doped (red open squares) coronene. Right panel: Zoom at the dominant C\,$1s$ excitation right after the excitation onset at 284.4\,eV for undoped coronene measured with higher energy resolution (85\,meV). The inset shows the C\,1$s$ core-level spectrum for an undoped coronene film grown on SiO$_2$ measured using x-ray photoemission spectroscopy (XPS).} \label{f2}
\end{figure}

In the left panel of Fig.\,\ref{f2} we show C\,$1s$ and K\,$2p$ core level excitations of undoped and potassium doped coronene. These data can be used to analyze the doping induced changes and furthermore to determine the stoichiometry of the potassium doped coronene films.  Moreover, the C\,$1s$ excitations represent transitions into empty C $2p$-derived levels, and thus allow to probe the projected unoccupied electronic density of states of carbon-based materials.\cite{Roth2008,Knupfer1999,Knupfer1995} Both spectra were normalized at the step-like structure in the region between 291\,eV and 293\,eV, i.\,e. to the $\sigma^*$ derived intensity, which is proportional to the number of carbon atoms. For the undoped case (black circles), we can clearly identify a sharp and strong feature in the range between 284 - 286\,eV, and some additional small features at 286.1\,eV, 286.9\,eV, 288.6\,ev and 290.2\,eV as well as a broad excitation at $\sim$ 294\,eV. Below $\sim$ 291\,eV the structures can be assigned to transitions into $\pi^*$
states representing the unoccupied electronic states. Also, because of the higher symmetry of the coronene molecules, compared to other carbon based materials like picene, we expect a degeneracy of the higher molecular orbitals which can be directly seen by the well separated features above 286\,eV. Such a well pronounced structure representing higher lying molecular orbitals is very similar to what was observed for fullerene. \cite{Chen1991,Wastberg1994,Sohmen1992,Knupfer1995} 
When we focus at the dominating excitation feature right after the excitation onset, as shown in the right panel of Fig.\,\ref{f2}, we can identify a characteristic fine structure with maxima at 284.75\,eV, 284.85\,eV, 285.15\,eV and 285.35\,eV. These features can be identified with maxima in the unoccupied density of states, since the carbon 1$s$ levels of the different C atoms in coronene are virtually equivalent as revealed by x-ray photoemission spectra (cf. Fig.\,\ref{f2} and Ref.(\cite{Schroeder2002})). The peak width of the C\,1$s$ photoemission line is smaller than 1\,eV as seen in the inset of Fig.\,\ref{f2} (Notice: The energy resolution for the XPS-measurements is $\approx$ 0.35\,eV. The broadening of the spectral linewidth is a result of lifetime-effects, very similar to what was observed for C$_{60}$, where all carbon atoms are symmetrically equivalent\cite{POIRIER1991}). The observation of four structures in Fig.\,\ref{f2} (right panel) is in very good agreement with first principle band structure calculations for undoped coronene which predict four close lying conduction bands (arising from the doubly degenerate LUMO  with e$_{1g}$ symmetry as well as the doubly degenerate LUMO+1 with e$_{2u}$ symmetry) in this energy region.\cite{Kosugi2011} Interestingly, a related fine structure was reported for picene both experimentally \cite{Roth2010} and theoretically \cite{Kosugi2009} which shows also superconductivity upon doping. The step-like structure above 291\,eV corresponds to the onset of transitions into $\sigma^*$-derived unoccupied levels.

\par

Also in the case of potassium doped coronene (red open squares) the spectrum is dominated by a sharp excitation feature in the range between 284 - 286\,eV and, in addition, by K\,$2p$ core excitations, which can be observed at 297.2\,eV and 300\,eV, and which can be seen as a first evidence of the successful doping of the sample. In particular, the well structured K\,2$p$ core excitations signal the presence of K$^+$ ions, in agreement with other studies of potassium doped molecular films.\cite{Knupfer2001,Flatz2007,Roth2011} Their spectral shape is clearly different from the much broader and less structured K\,2$p$ core excitation spectrum of a pure potassium multiplayer.\cite{Ma1992} We observe a clear broadening of the first C\,$1s$ feature which might arise from a change of the binding energy of the C\,1$s$ core levels because of the introduced potassium atoms as well as lifetime effects in the metallic doped coronene. Moreover, upon charging the coronene molecules will most likely undergo a Jahn-Teller distortion \cite{Sato2003}, which leads to a splitting of the electronic molecular levels and thus a spectral broadening in our data. Also, band structure calculations of K$_3$coronene \cite{Kubozono2011} predict shifts of the conduction bands as compared to pristine coronene which would result in a broadening as seen in Fig.\,\ref{f2}. Additionally, also a mixing of different phases, which we can not exclude, can result in a broadening of the C\,1$s$ signal.

Importantly, a clear reduction of the spectral weight of the first C\,$1s$ excitation feature is observed in Fig.\,\ref{f2} upon doping. Taking into account the four conduction bands that contribute to this intensity in the undoped case, this reduction is a clear signal for the successful doping. The stoichiometry analysis can further be substantiated by comparing the K\,$2p$ and C\,$1s$ core excitation intensities in comparison to other doped molecular films with known stoichiometry such as K$_6$C$_{60}$ \cite{Knupfer2001} like in previous publications.\cite{Flatz2007,Roth2008,Roth2011} The results shown in the left panel of Fig.\,\ref{f2} indicate a doping level of K$_{2.8}$coronene, which again is in very good agreement to the other results discussed above.

\par

\begin{figure}[h]
\centering
\includegraphics[height=5.9cm]{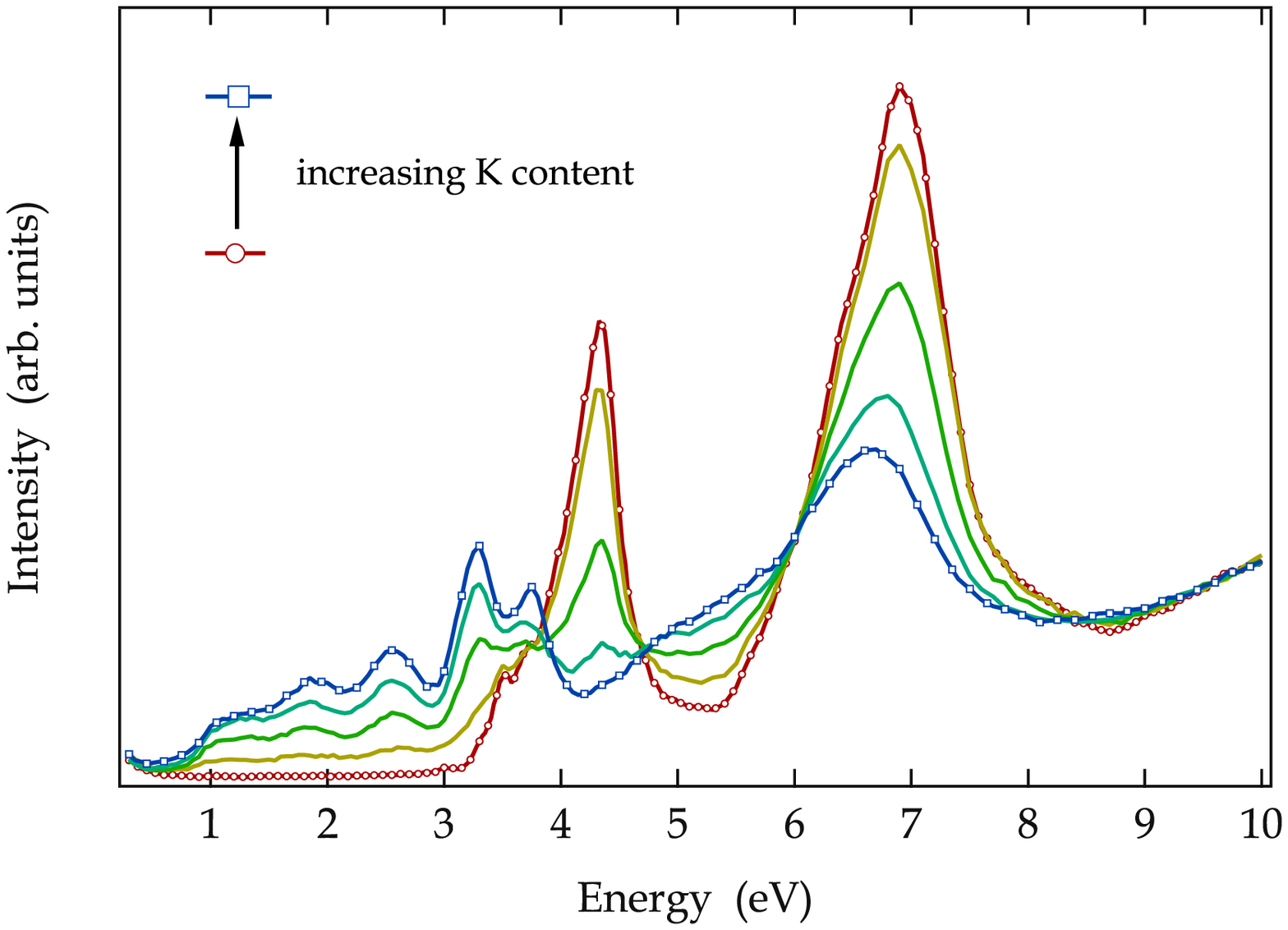}
\includegraphics[height=5.9cm]{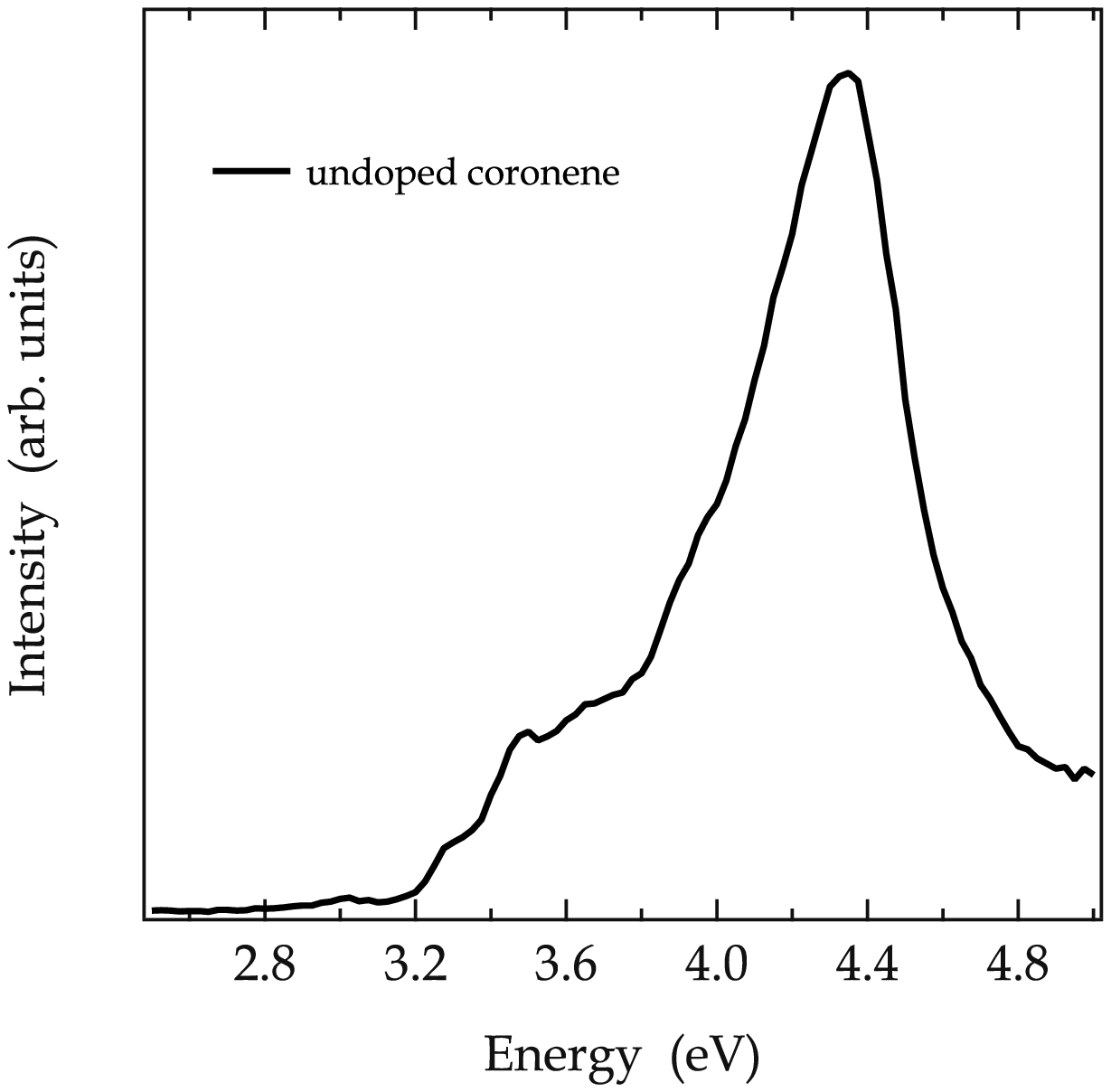}
\caption{Left panel: Evolution of the loss function of coronene in the range of 0 - 10\,eV upon potassium doping. All spectra were normalized in the high energy region between 9 - 10\,eV. (K content increases from bottom (red open circles) to top (blue open squares))  Right panel:  Loss function of solid coronene measured with a momentum transfer of q = 0.1\,\AA$^{-1}$ at 20\,K.} \label{f3}
\end{figure}

Doping of coronene also causes major changes in the electronic excitation spectrum as revealed in Fig.\,\ref{f3}, where we show a comparison of the loss functions in an energy range of 0-10\,eV measured using EELS for different doping steps. These data are taken with a small momentum transfer $\textbf{q}$ of 0.1\AA$^{-1}$, which represents the optical limit. For undoped coronene, we can clearly identify two main  maxima at about 4.3\,eV and 6.9\,eV, which are due to excitations between the occupied and unoccupied electronic levels. In addition, zooming into the energy region around the excitation onset in the experimental spectra reveals an optical gap of 2.8\,eV (cf.\,Fig.\,\ref{f3}). This onset also represents a lower limit for the band gap of solid coronene. The excitation onset of coronene is followed by five additional well separated features at 3\,eV, 3.3\,eV, 3.5\,eV, 3.7\,eV and 3.95\,eV. The main features of our spectrum are in good agreement with previous EEL measurements in the gas phase \cite{Khakoo1990,Abouaf2009} and optical absorption data.\cite{Nijegorodov2001,Ohno1972}

\par

In general, the lowest electronic excitations in organic molecular solids usually are excitons, i.\,e. bound electron-hole pairs.
\cite{Pope1999,Knupfer2003,Lof1992,Hill2000} The decision criterion that has to be considered in order to analyse the excitonic
character and binding energy of an excitation is the energy of the excitation with respect to the so-called transport energy gap, which
represents the energy needed to create an unbound, independent electron-hole pair. Different values from 3.29\,eV \cite{Schroeder2002} up to 3.54\,eV \cite{Rieger2008} and 3.62\,eV \cite{Djurovich2009} for coronene are published in previous publications. Consequently, the lowest excitation that is observed can safely be attributed to a singlet exciton.

\par

Upon doping, the spectral features become broader, and a downshift of the second, major excitation can be observed. We assign this downshift to a relaxation of the molecular structure of coronene as a consequence of the filling of anti-bonding $\pi^*$ levels. Furthermore, a decrease of intensity of the feature at 4.3\,eV upon potassium intercalation is visible. In addition, for the doped films new structures at 1.15\,eV, 1.9\,eV, 2.55\,eV, 3.3\,eV and 3.75\,eV are observed in the former gap of coronene. Taking into account the structural relaxation upon doping and the fact that the doped molecules are susceptible to a Jahn-Teller distortion\cite{Sato2003}, one would expect additional excitations in this energy region similar to what has been observed previously for other doped $\pi$-conjucated material.\cite{Lane1996,Flatz2007,Golden1995} These then arise from excitations between the split former HOMO and LUMO of coronene and excitations from the former LUMO to LUMO+1, which become possible as soon as the LUMO is occupied. A direct assignment of the observed features however requires further investigations.

\par

\begin{figure}[h]
\centering
\includegraphics[width=0.6\textwidth]{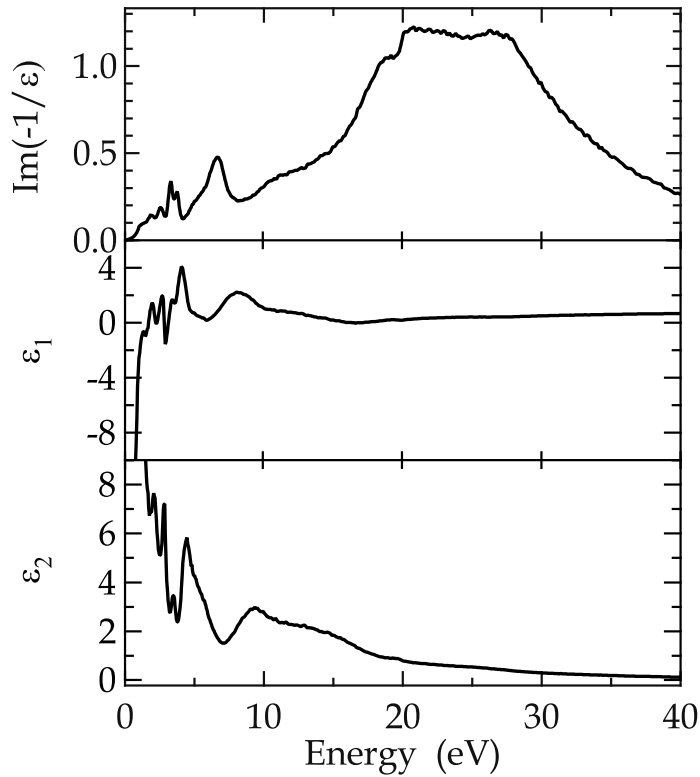}
\caption{Loss function (Im(-1/$\epsilon$), real part ($\epsilon_1$) and imaginary part ($\epsilon_2$) of the dielectric function of K doped coronene. The momentum transfer is $q$\,=\,0.1\,\AA$^{-1}$. Note that in contrast to Fig.\,\ref{f3}, the loss function is corrected for the contribution of the direct beam and multiple scattering.} \label{f4}
\end{figure}

In order to obtain deeper insight into potassium introduced variations, we have analyzed the measured loss function, Im(-1/$\epsilon$), of doped coronene using a Kramers Kronig analysis (KKA).\cite{Fink1989} This analysis has been carried out for a metallic ground state since the observation of superconductivity \cite{Kubozono2011} as well as band structure calculations \cite{Kosugi2011} signal such a ground state for the stoichiometry of K$_3$coronene. Furthermore, the evolution of the loss function in Fig.\,\ref{f3} (left panel) indicates a filling of the former energy gap.

\begin{figure}[h]
\centering
\includegraphics[width=0.6\textwidth]{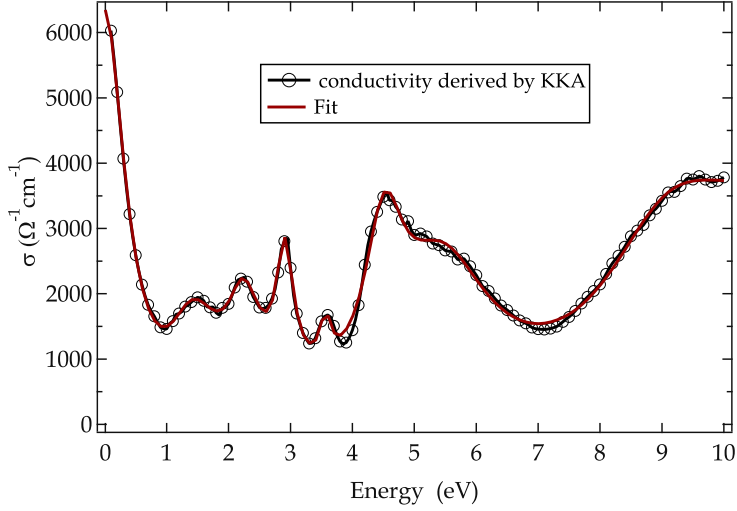}
\caption{Optical conductivity $\sigma = \omega\epsilon_0\epsilon_2$ of K doped coronene for a momentum transfer of
$q$\,=\,0.1\,\AA$^{-1}$ (black circles) derived by a Kramers-Kronig analysis of the EELS intensity. Additionally, the result of a Drude-Lorentz fit is shown (red line).} \label{f5}
\end{figure}

In Fig.\,\ref{f4} we present the results of this analysis in a wide energy range between 0 - 40\,eV. The loss function (cf. Fig.\,\ref{f4} upper panel) is dominated by a broad maximum in the range between 20 - 27\,eV which can be assigned to the $\pi + \sigma$ plasmon, a collective excitation of all valence electrons in the system. Various interband excitations at 2 - 20\,eV can be observed as maxima in the imaginary part of the dielectric function, $\epsilon_2$ (lower panel).
Most interestingly, $\epsilon_2$ shows in the energy range between 0 - 6\,eV only 4 maxima in contrast to the 5 features in the loss function. The zero crossing, which can be seen as a definition of a charge carrier plasmon, near 1.8\,eV in the real part of the dielectric function, $\epsilon_1$ (middel panel) as well as the absence of a peak at the same energy in the imaginary part of the dielectric function and accordingly the optical conductivity, $\sigma$, (cf. Fig.\,\ref{f5}), leads to the conclusion that the second spectral feature at 1.9\,eV represents a collective excitation (density oscillation), and we assign it to the charge carrier plasmon of doped coronene.

To test the consistency of our KKA analysis we check the sum rules as described elsewhere.\cite{Mahan2000}
An evaluation of this sum rule for the loss function and the dielectric function after our KKA results in a very good agreement of the two
values and further with the value what is expected from a calculation of the electron density of doped coronene. Moreover, a further sum rule which is only valide for metallic systems can be employed \cite{Mahan2000}:

\begin{align*}
 \int\limits_{0}^{\infty} d\omega \frac{\operatorname{Im}\left(-\frac{1}{\epsilon(\textbf{q},\omega)}\right)}{\omega} = \frac{\pi}{2}.
\end{align*}

Here, after our KKA we arrive at a value of 1.562, very close to the expectation of $\frac{\pi}{2}$.

\par

In order to obtain access to further information we show in Fig.\,\ref{f5} the optical conductivity, $\sigma$, of potassium doped coronene. As we can see from Fig.\,\ref{f5} the optical conductivity consists of a free electron contribution at low energies due to intraband transitions in the conduction bands and some additional interband contributions. From the optical conductivity at $\omega = 0$ we can derive the $dc$-conductivity of K$_{2.8}$coronene of about 6300\,$\Omega^{-1}\mathrm{cm}^{-1}$. We have additionally fitted the optical conductivity using a simple Drude-Lorentz model:

\begin{align}
\epsilon(\omega) = 1 - \frac{\omega_D^2}{\omega^2+i\gamma_D\omega} + \sum_{j} \frac{f_j}{\omega_{j_o}^2-\omega^2-i\gamma_j\omega}
\end{align}

Within this model, the free electron contribution is described by the Drude part ($\omega_D$,$\gamma_D$) and the interband transitions are represented by the Lorentz oscillators ($f_j$, $\gamma_j$ and $\omega_{j_o}$). The resulting fit parameters are given in Table\,\ref{tab1} and are also shown in Fig.\,\ref{f5} as red line. This Figure demonstrates that our model description of the data is very good. We note, that the result of our fit also describes $\epsilon_1$ very well, which demonstrates the consistency of our description.

\begin{table}[h]
  \centering
\begin{tabular}{c| c c c || c c}
    \hline
  i \quad & \quad $\omega_{j_o}$ (eV) & \quad  $\gamma_j$ (eV) & \quad  $f_j$ (eV) &\quad  $\gamma_D$ (eV) & \quad  $\omega_D$ (eV)\\
  \hline
  1 \quad &\quad  1.47 & \quad  1.01 & \quad  3.12 &\quad  0.40 & \quad  4.35\\
  2 \quad &\quad  2.25 & \quad  0.65 & \quad  2.55 & & \\
  3 \quad &\quad  2.90 & \quad  0.34 & \quad  2.22 & & \\
  4 \quad &\quad  3.57 & \quad  0.22 & \quad  1.06 & & \\
  5 \quad &\quad  4.52 & \quad  0.77 & \quad  3.67 & & \\
  6 \quad &\quad  5.48 & \quad  1.83 & \quad  5.26 & & \\
  7 \quad &\quad  9.15 & \quad  2.75 & \quad  7.00 & & \\
  8 \quad &\quad  10.64 & \quad  2.84 & \quad  6.76 & & \\
\hline
 \end{tabular}
\caption{Parameters derived from a Drude-Lorentz fit of the optical conductivity (as shown in Fig.\,\ref{f5}) using formula (1). The Drude
part is given by the plasma energy $\omega_D$ and the width of the plasma (damping) $\gamma_D$, while $f_j$, $\gamma_j$ and $\omega_{j_o}$ are the oscillator strength, the width and the energy position of the Lorentz oscillators.} \label{tab1}
\end{table}

We arrive at an unscreened plasma frequency $\omega_D$ of about 4.35\,eV. Furthermore, we can compare this value with the plasma frequency which we expect if we take the additional six conduction electrons per unit cell (2 coronene molecules per unit cell) in K$_3$coronene into account. We arrive at a value of 3.42\,eV which is somewhat lower than what we have derived using our fit procedure. This might be related to an effective mass of the charge carriers in doped coronene, which is reduced as compared to the free electron value, $m_0$. For related (undoped) organic crystals such as rubrene \cite{Machida2010}, PTCDA \cite{Temirov2006,Ueno2008} or pentacene \cite{Doi2005} an effective mass also lower than $m_0$ has been deduced previously. Finally, a comparison of the unscreened and screened plasma frequencies can be used to derieved the averaged screening $\epsilon_{\infty}$ of the charge carrier plasmon \cite{Roth2011}. This would give a value of $\epsilon_{\infty} \sim (\nicefrac{4.35}{1.8})^2 \sim 5.8$. We note however that this is a very rough approximation for K$_3$coronene, since there are close lying interband excitations in the corresponding energy region. 

\section{summary}

To summarize, we have investigated the electronic properties of potassium doped coronene compared to undoped coronene using electron energy-loss spectroscopy in transmission. Core level excitation data signal the formation of a doped phase with a stoichiometry close to K$_3$coronene, which is reported to be superconductive. The reduction of the lowest lying C\,$1s$ excitation features clearly demonstrates that potassium addition leads to a filling of the coronene conduction bands. Furthermore, the electronic excitation spectrum changes substantially upon doping. In particular, several new low energy features show up upon potassium intercalation, and one of these features can be associated with the charge carrier plasmon.

\begin{acknowledgments}
We thank R. Sch\"onfelder, R. H\"ubel and S. Leger for technical assistance. This work has been supported by the Deutsche Forschungsgemeinschaft (grant number KN393/14).
\end{acknowledgments}

\end{document}